\documentclass[12pt,a4paper]{article}

\usepackage{subfigure}
\usepackage{array}
\usepackage{graphicx}
\usepackage{amsmath}
\usepackage{amssymb}
\usepackage{mathrsfs}
\usepackage{bm}
\usepackage{color}
\usepackage{slashed}
\usepackage{threeparttable}
\usepackage[colorlinks, linkcolor=black, anchorcolor=black, citecolor=black]{hyperref}
\usepackage[amsmath,thmmarks,hyperref]{ntheorem}
\usepackage{soul}

\title{Annihilation rate of $2^{-+}$ charmonium and bottomonium}
\author{Tianhong Wang\footnote{thwang.hit@gmail.com}, Guo-Li Wang\footnote{gl\_wang@hit.edu.cn}, Wan-Li Ju, Yue Jiang\\
{\it \small   Department of Physics, Harbin Institute of Technology,
Harbin, 150001, China} }
\date{\today}

\begin{document}
\maketitle

\begin{abstract}
The $1^1D_2~(c\bar c)$ state is the ground state of spin-singlet D-wave charmonia. Although it has not been found yet, the experimental data accumulate rapidly. This charmonium
attracts more and more attention, especially when the BaBar Collaboration finds that
the $X(3872)$ particle has negative parity. In this paper we calculate the
double-gamma and double-gluon annihilation processes of $^1D_2$ charmonia and bottomonia by using the instantaneous Bethe-Salpeter
method. We find the relativistic corrections make the decay widths
of $1^1D_2~(c\bar c)$ 2$\sim$5 times smaller than the
non-relativistic results. If this state is below the
$D^0{D^{0}}^\ast$ threshold, we can use the sum of annihilation
widths and EM transition widths to estimate the total decay width.
Our result for $1{^1D_2}~(c\bar c)$ with $m=3820$ GeV is $\Gamma=
432$ keV. The dominant decay channel $1^1D_2~(c\bar c)\to h_c \gamma$,
whose branching ratio is about 90\%, can be used to discover this state.
\end{abstract}

\section{Introduction}

The $2^{-+}$ state draws more and more
attention~\cite{Chao1,Jia,Li,Kalash,Burns,Lange,Chao2,wang2} recently. One main
reason is that the latest result of the BaBar
Collaboration~\cite{Babar} favors negative parity for $X(3872)$, which has induced many discussions about the possibility of assigning the
$1{^1D_2}~(c\bar c)$ state to this particle. However, theoretical
calculations show that the $2^{-+}$ assignment strongly
contradicts with the experimental data of the electromagnetic decays of
$X(3872)$~\cite{Jia, Li,Kalash,wang2}. It's very
interesting to study the properties of the $1{^1D_2}~(c\bar
c)$ state, which is the only ground state of spin-singlet
D-wave charmonia. The discovery of this particle will greatly support the quark potential models and be helpful in understanding the non-perturbative properties of QCD.

These models predict the mass region of this particle is $3760\sim 3840$
MeV~\cite{Jia}, below the threshold value of
$D^0{D^0}^\ast$. This means no OZI decay channels are allowed.
So the electromagnetic and light hadronic decay channels are
important for the discovery of this particle. For example, the clean diphoton decay channel will play an important role in determining the
inner structure of these particles. This channel can be used to distinguish
mesons with a $q\bar q$ structure from those without~\cite{AB}. We note that non-relativistic calculations of the two-gluon annihilation process of D-wave mesons have been
performed recently in Refs.~\cite{Chao1,ELQ,GB}. All of them get a relatively large result. In Ref.~\cite{Chao1}, Chao {\sl et al.} have calculated this process to the third
order of $\alpha_s$ within NRQCD formalism. They show the next-to-leading order QCD corrections contribute enhancement factor of 1.8 and 1.5 for charmonium and bottomonium $1^1D_2$ states, respectively.
With the non-relativistic approximation, the decay widths of D-wave mesons are related to the
second derivative of the radial wave functions at the origin. This
method will result in large errors, since the full behavior of the
wave function (or the relativistic correction) is significant for D-wave mesons.

The semi-relativistic calculation~\cite{AB} and the relativistic calculation~\cite{Munz} have
been performed before. In Ref.~\cite{AB}, the transition amplitude is calculated based on the non-relativistic wave function and free quark-antiquark annihilation Feynman diagrams, and the dependence of the meson mass is introduced by adding an additional term in the Lagrangian. Ref.~\cite{Munz} gives a relativistic result, but there they use a potential with the timelike vector spin structure and expend the wave function (amplitude) in a set of Laguerre basis functions.
Since large relativistic correction is expected for D-wave meson, a careful relativistic calculation with the wave function given by a more reasonable way is very necessary. This will be helpful for the discovery and study of this particle for the future experiments.

In the previous papers~\cite{wang3,wang4,wang5}, we have calculated the two-photon and two-gluon annihilation rates of $0^{-+}$, $0^{++}$ and $2^{++}$ states by using the relativistic Salpeter method~\cite{BS1,BS2}. Large relativistic corrections are found, especially for the two P-wave states. There we made an approximation that the time components of the quark-antiquark momenta inside a meson are constants, setting $p_1^0=p_2^0=\frac{M}{2}$ (Refs.~\cite{AB} use the similar approximation). In this paper, we study the annihilation processes of $2^{-+}$ states within the same formalism, but with some improvements. Except for the method we used before, we also perform the calculation without that approximation and compare the results of these two methods. We examine the accuracy of this approximation and provide some useful information for future study.

The paper is organized as follows: In section~2, we present the general form of the $2^{-+}$ wave function and the coupling Bethe-Salpeter (BS) equations fulfilled by this wave function. We also give the transition amplitude for the two-photon (gluon) annihilation processes within Mandelstam formalism. In section 3 we show and discuss the numerical results.

\section{Theoretical calculations}

The general wave function of the $2^{-+}$ state with mass $M$, momentum $P$ and polariztion tensor $\epsilon_{\mu\nu}$ can be written as~\cite{wang2}
\begin{equation}
\label{wf}
\begin{aligned}
\varphi_{2^{-+}}(q_\perp)&=\epsilon_{\mu\nu}q_\perp^\mu
q_\perp^\nu[f_1(q_\perp)+\frac{\slashed{P}}{M}f_2(q_\perp)
+\frac{\slashed{q}_\perp}{M}f_3(q_\perp)+\frac{\slashed{P}\slashed{q}_\perp}{M^2}f_4(q_\perp)]\gamma^5,
\end{aligned}
\end{equation}
where $q_\perp^\mu\equiv q^\mu-\frac{P\cdot q}{M^2}P^\mu$; $q$ is the relative momentum between the constituent quark and antiquark. The constraint condition (see the last one in Eq.~(\ref{salpeter})) of the scalar functions $f_i$ has the following form:
\begin{equation}
\label{wfcons}
\begin{aligned}
&f_3(q_\perp)=\frac{f_1(q_\perp)M(m_1\omega_2-m_2\omega_1)}{q_\perp^{2}(\omega_1+\omega_2)},\\
&f_4(q_\perp)=\frac{-f_2(q_\perp)M(\omega_1+\omega_2)}{(m_1\omega_2+\omega_1m_2)},
\end{aligned}
\end{equation}
where $\omega_i=\sqrt{m_i^2-q_\perp^2}=\sqrt{m_i^2+\vec q^2}$. In the equal mass case, we get $\omega_1=\omega_2$ and $f_3=0$.

With the same method used in Ref.~\cite{wang1}, we can get the
coupled instantaneous BS equations for the $2^{-+}$ state
\begin{equation}
\label{coupleeq}
\begin{aligned}
&(M-2\omega_1)(f_1(\vec q)+\frac{\omega_1}{m_1}f_2(\vec q))\\
&=-\int d^3\vec k\frac{3}{2\vec q^4m_1\omega_1}[(\vec q\cdot\vec
k)^2- \frac{1}{3}\vec q^2 \vec k^2][m_1(V_s-V_v)(m_1f_2(\vec
k)+\omega_1f_1(\vec k))-(V_s+V_v)\vec k\cdot\vec q f_2(\vec k)],\\
&(M+2\omega_1)(f_1(\vec q)-\frac{\omega_1}{m_1}f_2(\vec q))\\
&=-\int d^3\vec k\frac{3}{2\vec q^4m_1\omega_1}[(\vec q\cdot\vec
k)^2- \frac{1}{3}\vec q^2 \vec k^2][m_1(V_s-V_v)(m_1f_2(\vec
k)-\omega_1f_1(\vec k))-(V_s+V_v)\vec k\cdot\vec q f_2(\vec k)].
\end{aligned}
\end{equation}

By solving above equations, we can get the mass spectrum and corresponding wave functions. $f_1$ and $f_2$ fulfill the normalization condition:
\begin{equation}
\begin{aligned}\label{eq:norm2}
\int \frac{d^3\vec q}{(2\pi)^3} \frac{16}{3}f_1f_2\frac{\omega_1}{m_1}\vec q^4 = 10 M.
\end{aligned}
\end{equation}

According to Mandelstam formalism~\cite{Man}, the relativistic transition amplitude for the double-photon decay processes (see Fig.~1) can be written as
\begin{figure}\label{feyn}
\centering
\subfigure[]{\includegraphics[scale=0.8]{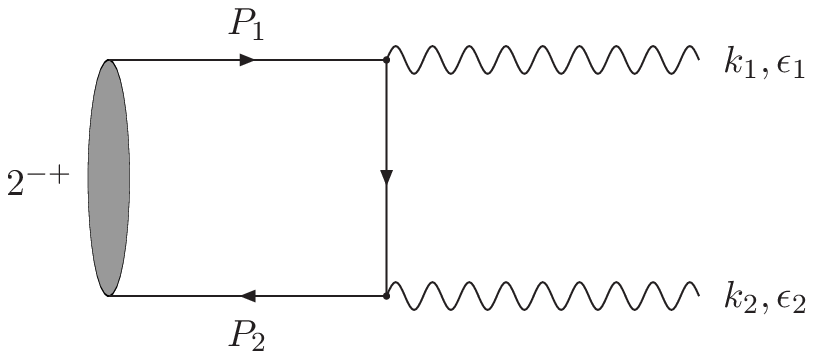}}
\hspace{5 mm}
\subfigure[]{\includegraphics[scale=0.8]{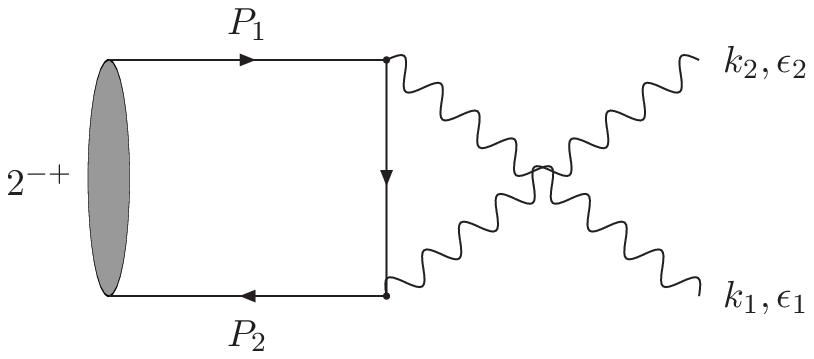}}
\caption{Feynman diagrams for $2^{-+}\rightarrow\gamma\gamma$.}
\end{figure}

\begin{figure}\label{wf}
\centering
\includegraphics[scale=0.5]{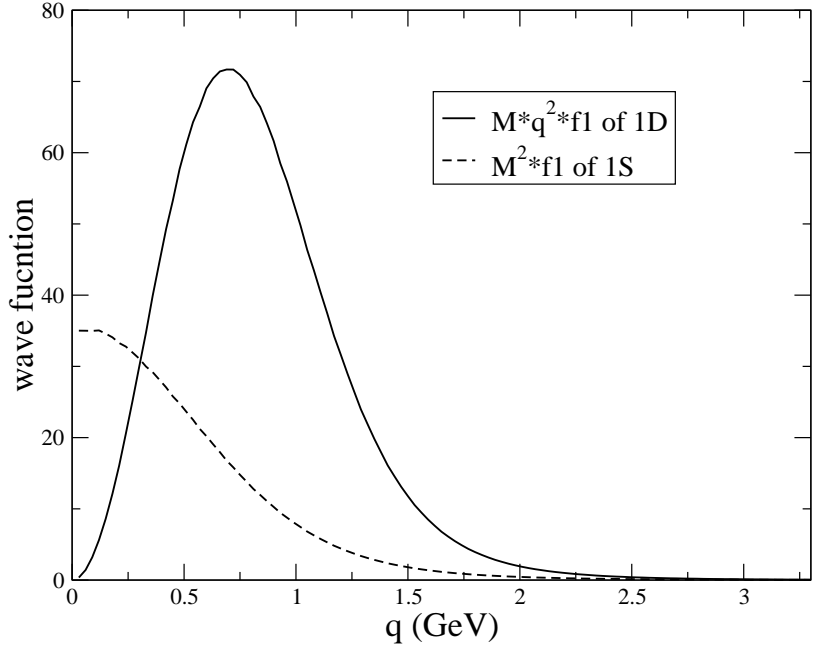}
\caption{Wave functions of $1^1D_2(c\bar c)$ and $\eta_c~(1^1S_0)$.}
\end{figure}

\begin{equation}
\label{feynmatrix}
\begin{aligned}
{\mathcal M}&=\sqrt{3}(ee_q)^2\int\frac{d^4q}{(2\pi)^4}{\rm Tr}\{\chi(q)[\slashed\epsilon_2\frac{1}{\slashed p_1-\slashed k_1-(m_1-i\epsilon)}\slashed\epsilon_1+\slashed\epsilon_1\frac{1}{\slashed p_1-\slashed k_2-(m_1-i\epsilon)}\slashed\epsilon_2]\}\\
&={\mathcal M_1}+{\mathcal M_2},
\end{aligned}
\end{equation}
where $\chi(q)$ is the BS wave function of the initial meson; $k_i$ and $\epsilon_i$ are momenta and polarization vectors of final photons, respectively; $e_q=\frac{2}{3}$ for charmonium and $e_q=-\frac{1}{3}$ for bottomonium.

In the following, we give the detailed calculation for $\mathcal M_1$ in Eq.~(\ref{feynmatrix}).
\begin{equation}
\label{m1}
\begin{aligned}
{\mathcal M_1}&=\sqrt{3}(ee_q)^2\int\frac{d^4q}{(2\pi)^4}{\rm Tr}\{\chi(q)\slashed\epsilon_2\frac{1}{\slashed p_1-\slashed k_1-(m_1-i\epsilon)}\slashed\epsilon_1\}\\
&=\sqrt{3}(ee_q)^2\int\frac{d^4q}{(2\pi)^4}{\rm Tr}\{\chi(q)\frac{\slashed\epsilon_2(\alpha_1\slashed P+\slashed q-\slashed k_1+m_1)\slashed\epsilon_1}{(\alpha_1P+q)^2-2(\alpha_1P+q)\cdot k_1-m_1^2+i\epsilon}\}\\
&=\sqrt{3}(ee_q)^2\int\frac{d^4q}{(2\pi)^4}{\rm Tr}\{\chi(q)\frac{\slashed\epsilon_2((\alpha_1+\frac{q_P}{M})\slashed P+\slashed q_\perp-\slashed k_1+m_1)\slashed\epsilon_1}{q_P^2-A^2+i\epsilon}\}.
\end{aligned}
\end{equation}
In the second line we have used $p_1^\mu=\alpha_1P^\mu+q^\mu$, where $\alpha_1=\frac{m_1}{m_1+m_2}$; in the third line we have used $q^\mu = q_\parallel^\mu+q_\perp^\mu$ and $q_\parallel^\mu = \frac{q_P}{M}P^\mu$, where $q_P=\frac{P\cdot q}{M}$. We also define
\begin{equation}
\label{Asquare}
A^2=-(q_\perp^2-\frac{M^2}{4}-2q_\perp\cdot k_1-m_1^2)=(\vec q-\vec k_1)^2+m_1^2.
\end{equation}

The BS wave function $\chi(q)$, which is related to Eq.~(\ref{wf}) through
\begin{equation}
\label{bswf}
\varphi(q_\perp^\mu) \equiv i \int\frac{dq_p}{2\pi}\chi(q_\parallel^\mu, q_\perp^\mu),
\end{equation}
has the form~\cite{wang1}:
\begin{equation}
\label{bswfdef}
\begin{aligned}
\chi(q)=S_1(p_1)\eta(q_\perp)S_2(p_2),
\end{aligned}
\end{equation}
where
\begin{equation}
\label{eta}
\eta(q_\perp)=\int d^3k_\perp V(q_\perp,k_\perp)\varphi(k_\perp).
\end{equation}

$S_i(p_i)$ is defined as
\begin{equation}
\label{propagator}
\begin{aligned}
&S_1(p_1)=\frac{\Lambda^+_1}{q_P+\alpha_1M-\omega_1+i\epsilon}+\frac{\Lambda_1^-}{q_P+\alpha_1M+\omega_1-i\epsilon},\\
&S_2(p_2)=\frac{\Lambda^+_2}{-q_P+\alpha_2M-\omega_2+i\epsilon}+\frac{\Lambda_2^-}{-q_P+\alpha_2M+\omega_2-i\epsilon},\\
\end{aligned}
\end{equation}
where
\begin{equation}
\label{projector}
\begin{aligned}
&\Lambda_1^{\pm}=\frac{1}{2\omega_1}[\frac{\slashed P}{M}\omega_1\pm(\slashed q_\perp + m_1)],\\
&\Lambda_2^{\pm}=\frac{1}{2\omega_2}[\frac{\slashed P}{M}\omega_2\mp(\slashed q_\perp + m_2)].
\end{aligned}
\end{equation}
When $m_1=m_2$, we have $\omega_1=\omega_2$ and $\Lambda_1^{\pm}=\Lambda_2^{\mp}$.

Inserting Eq.~(\ref{bswfdef}) and Eq.~(\ref{propagator}) into Eq.~(\ref{m1}) and performing the counter integral over $q_P$ by considering~\cite{wang1}
\begin{equation}
\label{salpeter}
\begin{aligned}
&(M-\omega_1-\omega_2)\varphi^{++}(q_\perp) = \Lambda_1^+(q_\perp)\eta(q_\perp)\Lambda_2^+(q_\perp),\\
&(M+\omega_1+\omega_2)\varphi^{--}(q_\perp) = - \Lambda_1^-(q_\perp)\eta(q_\perp)\Lambda_2^-(q_\perp),\\ &\varphi^{+-}(q_\perp) = \varphi^{-+}(q_\perp) = 0,
\end{aligned}
\end{equation}
we get
\begin{equation}
\label{m11}
\begin{aligned}
{\mathcal M_1}=&i\sqrt{3}(ee_q)^2\int\frac{d^3q_\perp}{(2\pi)^3}\{C_1{\rm Tr}[\varphi^{++}\slashed\epsilon_2((1-\frac{\omega_1}{M})\slashed P+\slashed q_\perp-\slashed k_1+m_1)\slashed \epsilon_1]\\
&+C_2{\rm Tr}[\varphi^{++}\slashed\epsilon_2((\alpha_1-\frac{A}{M})\slashed P+\slashed q_\perp-\slashed k_1+m_1)\slashed\epsilon_1]\\
&+C_3{\rm Tr}[\varphi^{--}\slashed\epsilon_2(-\frac{\omega_1}{M}\slashed P+\slashed q_\perp-\slashed k_1+m_1)\slashed\epsilon_1]\\
&+C_4{\rm Tr}[\varphi^{--}\slashed\epsilon_2((\alpha_1-\frac{A}{M})\slashed P+\slashed q_\perp-\slashed k_1+m_1)\slashed\epsilon_1]\\
&+C_5{\rm Tr}[\Lambda_1^+\eta(q_\perp)\Lambda_2^-\slashed\epsilon_2((\alpha_1-\frac{A}{M})\slashed P+\slashed q_\perp-\slashed k_1+m_1)\slashed\epsilon_1]\\
&+C_6{\rm Tr}[\Lambda_1^-\eta(q_\perp)\Lambda_2^+\slashed\epsilon_2((\alpha_1+\frac{A}{M})\slashed P+\slashed q_\perp-\slashed k_1+m_1)\slashed\epsilon_1]\},
\end{aligned}
\end{equation}
where
\begin{equation}
\begin{aligned}
&C_1=-\frac{1}{(\alpha_1M-\omega_1)^2-A^2}, ~~C_2=-\frac{M-2\omega_1}{2A[(\alpha_1M-\omega_1)^2-A^2]},\\
&C_3=-\frac{1}{(\alpha_1M+\omega_1)^2-A^2},~~C_4= \frac{M+2\omega_1}{2A[(\alpha_1M+\omega_1)^2-A^2]},\\
&C_5=C_6=\frac{1}{2A[(A+\omega_1)^2-(\alpha_1M)^2]}.
\end{aligned}
\end{equation}
$\varphi^{++}$ and $\varphi^{--}$ are the positive and negative parts of the wave function, respectively. In the following, we will just consider the contribution of $\varphi^{++}$. This term is dominant and other terms can all be ignored.

$\varphi^{++}$ has the following form:
\begin{equation}
\label{positive}
\begin{aligned}
&\varphi^{++}=\epsilon_{\mu\nu}q^\mu_\perp q^\nu_\perp(A_1+A_2\slashed P+A_3\slashed P\slashed q_\perp)\gamma_5,
\end{aligned}
\end{equation}
where
\begin{equation}
A_1 = \frac{1}{2}(f_1+\frac{\omega_1}{m_1}f_2),~~~
A_2 = \frac{m_1}{2M\omega_1}(f_1+\frac{\omega_1}{m_1}f_2),~~~
A_3 = -\frac{1}{2M\omega_1}(f_1+\frac{\omega_1}{m_1}f_2).
\end{equation}

Inserting Eq.~(\ref{positive}) into Eq.~(\ref{m11}) and finishing the traces, we get
\begin{equation}
\label{m1integral}
\begin{aligned}
{\mathcal M}_1&=-4\sqrt{3}(ee_q)^2\epsilon_{\alpha\beta\sigma\delta}\epsilon_1^\alpha\epsilon_2^\beta P^\delta k_1^\sigma\epsilon_{\mu\nu}\int\frac{d^3q_\perp}{(2\pi)^3}q_\perp^\mu q_\perp^\nu (C_1+C_2)A_2.
\end{aligned}
\end{equation}
The integral can be expressed as the combination of $g^{\mu\nu}$, $k_1$ and $P$. Through some calculations, we get
\begin{equation}
\label{m1final}
\begin{aligned}
{\mathcal M}_1&=\sqrt{3}(ee_q)^2\epsilon_{\alpha\beta\sigma\delta}\epsilon_1^\alpha\epsilon_2^\beta P^\delta\epsilon_{\mu\nu}k_1^\mu k_1^\nu k_1^\sigma E.
\end{aligned}
\end{equation}
$E$ is defined as
\begin{equation}
\label{es}
\begin{aligned}
&E= \int\frac{ d^3q_\perp}{(2\pi)^3} (C_1+C_2)A_2\frac{|\vec q|^2}{2|\vec k_1|^2}(3\cos^2\theta-1).
\end{aligned}
\end{equation}
where $\theta$ is the angle between $\vec q$ and $\vec k_1$.

With the same method, $\mathcal M_2$ can be expressed as
\begin{equation}
\label{m2}
\begin{aligned}
{\mathcal M}_2&=\sqrt{3}(ee_q)^2\epsilon_{\alpha\beta\sigma\delta}\epsilon_1^\alpha\epsilon_2^\beta P^\delta\epsilon_{\mu\nu}k_1^\mu k_1^\nu k_1^\sigma E^\prime,
\end{aligned}
\end{equation}
where the prime means that we have changed $k_1$, $\epsilon_1$ into $k_2$, $\epsilon_2$ in Eq.~(\ref{m1}).

By adding Eq.~(\ref{m1final}) and (\ref{m2}), we finally get the transition amplitude
\begin{equation}
\label{mfinal}
\begin{aligned}
{\mathcal M}= {\mathcal M_1} +{\mathcal M_2} =\sqrt{3}(ee_q)^2\epsilon_{\alpha\beta\sigma\delta}\epsilon_1^\alpha\epsilon_2^\beta P^\delta\epsilon_{\mu\nu}k_1^\mu k_1^\nu k_1^\sigma (E + E^\prime).
\end{aligned}
\end{equation}
With the completeness condition of the polarization tensor
\begin{equation}
\begin{aligned}\label{eq:polar}
\sum_{\lambda}  \epsilon_{\mu\nu}^{(\lambda)}\epsilon_{\mu_1\nu_1}^{\ast(\lambda)} &= \frac{1}{2}[(-g_{\mu\mu_1}+\frac{P_\mu P_{\mu_1}}{M^2})(-g_{\nu\nu_1}+\frac{P_\nu P_{\nu_1}}{M^2})+(-g_{\mu\nu_1}+\frac{P_\mu P_{\nu_1}}{M^2})\\
&\times(-g_{\nu\mu_1}+\frac{P_\nu P_{\mu_1}}{M^2})]-\frac{1}{3}(-g_{\mu\nu}+\frac{P_\mu P_{\nu}}{M^2})(-g_{\mu_1\nu_1}+\frac{P_{\mu_1} P_{\nu_1}}{M^2}),
\end{aligned}
\end{equation}
we get the unpolarized transition amplitude squared
\begin{equation}
\label{msquare}
\sum|{\mathcal M}|^2= \frac{3(ee_q)^4M^8}{48}(E + E^\prime)^2.
\end{equation}
The two-photon and two-gluon annihilation rates have the respective forms:
\begin{equation}
\label{width}
\begin{aligned}
&\Gamma_{2\gamma} = \frac{1}{2!\cdot5\cdot16\pi\cdot M} \sum|{\mathcal M}|^2,\\
&\Gamma_{2g} = \frac{2}{9}(\frac{g_s}{ee_q})^4\Gamma_{2\gamma}.
\end{aligned}
\end{equation}

\section{Results and discussions}

When solving the BS equation, we have used the Cornell potential
\begin{equation}
\begin{aligned}
\label{Cornell}
V(\vec{q})=V_s(\vec{q})
+\gamma_0\otimes\gamma^0V_v(\vec{q}),\\
V_{s}(\vec{q})
=-(\frac{\lambda}{\alpha}+V_0)\delta^{3}(\vec{q})
+\frac{\lambda}{\pi^{2}}\frac{1}{(\vec{q}^{2}+\alpha^{2})^{2}},\\
V_v(\vec{q})=-\frac{2}{3\pi^{2}}
\frac{\alpha_{s}(\vec{q})}{\vec{q}^{2}+\alpha^{2}},\\
\alpha_s(\vec{q})=\frac{12\pi}{27}
\frac{1}{{\rm{ln}}(a+\frac{\vec{q}^2}{\Lambda_{QCD}})},
\end{aligned}
\end{equation}
where the following values for the parameters are used: $a=e=2.7183$, $\alpha$ = 0.06
GeV, $\lambda$ = 0.21 ${\rm GeV}^2$, $m_c$ = 1.62 GeV, $m_b=4.96$ GeV, $\Lambda_{QCD}$ =
0.27 GeV (for $b\bar b$, $\Lambda_{QCD}$=0.20 GeV). $V_0$ can be fixed by fitting the mass spectrum. However, there is no experimental value available now. So we just take $M(1D(c\bar c)) = 3820$ MeV as an example which lies in the range of $3.760 \sim 3.840$ GeV predicted by quark potential models~\cite{Jia}.  As for $M(1D(b\bar b))$, we take 10.15 GeV which is the same as that in Ref.~\cite{GI}. By doing so, we get $V_0$ = -0.144 GeV for charmonium and -0.15 GeV for bottomonium.

We present masses of the first five states in Table~\ref{spectrum}. One notices that for charmonium only the ground state is below the $DD^\ast$ threshold, while for bottomonium the first two states are below the $BB^\ast$ threshold. For these states, we can use the sum of double-gluon, double-photon and E1 decay widths to estimate the total width.
\begin{table*}[htb]
\caption{The predicted masses (${\rm GeV}$) of $2^{-+}$ charmonia and bottomonia.}
\label{spectrum}
\setlength{\tabcolsep}{0.5cm}
\centering
\begin{tabular*}{\textwidth}{@{}@{\extracolsep{\fill}}|c|c|c|c|c|c|}
\hline
Meson&1D &2D& 3D&4D&5D \\ \hline
{\phantom{\Large{l}}}\raisebox{+.2cm}{\phantom{\Large{j}}}
$c\bar c$&3.820 &4.151&4.405 &4.611 &4.781\\ \hline
{\phantom{\Large{l}}}\raisebox{+.2cm}{\phantom{\Large{j}}}
$b\bar b$&10.15  &10.45 &10.70 &10.90 &11.08\\ \hline
\end{tabular*}
\end{table*}

\begin{table*}[htb]
\caption{Double-gamma decay widths (eV) of $2^{-+}$ charmonia and bottomonia. Results in parentheses are obtained by using the $p_1^0 =p_2^0= \frac{M}{2}$ approximation.}
\label{gamma}
\setlength{\tabcolsep}{0.1cm}
\centering
\begin{tabular*}{\textwidth}{@{}@{\extracolsep{\fill}}|c|c|c|c|c|c|}
\hline
Meson&1D &2D& 3D&4D&5D \\ \hline
{\phantom{\Large{l}}}\raisebox{+.2cm}{\phantom{\Large{j}}}
$c\bar c$&11.6 (14.8) &13.4 (18.7) &13.4 (19.4) &12.9 (18.9) &12.0 (17.9) \\ \hline
{\phantom{\Large{l}}}\raisebox{+.2cm}{\phantom{\Large{j}}}
$b\bar b$&0.0475 (0.0590) &0.0768 (0.0959) &0.0958 (0.120) &0.108 (0.135) &0.116 (0.146) \\ \hline
\end{tabular*}
\end{table*}

\begin{table*}[htb]
\caption{Double-gluon decay widths (keV) of $2^{-+}$ charmonia and bottomonia. Results in parentheses are obtained by using the $p_1^0 =p_2^0= \frac{M}{2}$ approximation.}
\label{gluon}
\setlength{\tabcolsep}{0.2cm}
\centering
\begin{tabular*}{\textwidth}{@{}@{\extracolsep{\fill}}|c|c|c|c|c|c|}
\hline
Meson&1D &2D& 3D&4D&5D \\ \hline
{\phantom{\Large{l}}}\raisebox{+.2cm}{\phantom{\Large{j}}}
$c\bar c$&35.6 (45.6) &41.3 (57.5) &41.4 (59.6) &39.6 (58.1) &37.1 (55.1) \\ \hline
{\phantom{\Large{l}}}\raisebox{+.2cm}{\phantom{\Large{j}}}
$b\bar b$&0.883 (1.20) &1.43 (1.78) &1.78 (2.23) &2.01 (2.52) &2.16 (2.71)  \\ \hline
\end{tabular*}
\end{table*}

\begin{table*}[!ht]
\caption{Double-gamma decay widths (eV) of $2^{-+}$ charmonia and bottomonia of different models. In parentheses, meson masses (GeV) of different models are presented.}
\label{compare1}
\setlength{\tabcolsep}{0.05cm}
\centering
\begin{tabular*}{\textwidth}{@{}@{\extracolsep{\fill}}|c|c|c|c|c|c|}
\hline
Model&Ours &Ref~\cite{AB}&Ref~\cite{Munz}&Ref~\cite{Crater}&Ref~\cite{Novikov}  \\ \hline
{\phantom{\Large{l}}}\raisebox{+.2cm}{\phantom{\Large{j}}}
$1D(c\bar c)$ &11.6 (3.820) &20 (3.84)&13.6 (3.84)&65.7$\sim$73.2 (3.872)&$\sim50$ ($3.5\sim3.7$) \\ \hline
{\phantom{\Large{l}}}\raisebox{+.2cm}{\phantom{\Large{j}}}
$2D(c\bar c)$ &13.4 (4.151)& 35 (4.21)&20.2 (4.28)& & \\ \hline
{\phantom{\Large{l}}}\raisebox{+.2cm}{\phantom{\Large{j}}}
$1D(b\bar b)$ &0.0475 (10.15)&0.033 (10.15)&0.0513 (10.13)& & \\ \hline
{\phantom{\Large{l}}}\raisebox{+.2cm}{\phantom{\Large{j}}}
$2D(b\bar b)$ &0.0768 (10.45)&0.069 (10.45)&0.0962 (10.47)&& \\ \hline
\end{tabular*}
\end{table*}

\begin{table*}[!ht]
\caption{Double-gluon decay widths (keV) of $2^{-+}$ charmonia and bottomonia of different models.  In parentheses, meson masses (GeV) of different models are presented.}
\label{compare2}
\setlength{\tabcolsep}{0.1cm}
\centering
\begin{threeparttable}
\begin{tabular*}{\textwidth}{@{}@{\extracolsep{\fill}}|c|c|c|c|c|c|c|}
\hline
Model&Ours &Ref~\cite{Chao1}\tnote{$\ast$}& Ref~\cite{ELQ}&Ref~\cite{Volk} &Ref~\cite{GB}&Ref~\cite{Novikov}\\ \hline
{\phantom{\Large{l}}}\raisebox{+.2cm}{\phantom{\Large{j}}}
$1D(c\bar c)$ &35.6&155  &110 (3.815) &60 &190 (3.837)&$\sim$60\\ \hline
{\phantom{\Large{l}}}\raisebox{+.2cm}{\phantom{\Large{j}}}
$1D(b\bar b)$ &0.883 &3.22 &&&& \\ \hline
\end{tabular*}
   \begin{tablenotes}
       \footnotesize
       \item[$\ast$] This is the leading order results. To the $\mathcal{O}(\alpha_s^3)$ order, the results is 274 keV and 4.70 keV for $c\bar c$ and $b\bar b$, respectively
   \end{tablenotes}
\end{threeparttable}
\end{table*}

Except for the above method, we also perform the calculation as we did in Ref~\cite{wang3,wang4,wang5}. That is, the $q^0$ dependence of the integrand in Eq.~(\ref{feynmatrix}) only comes from $\chi(q)$ with the condition  $p_1^0 = \frac{M}{2}$ (which is equivalent to $p_1=\alpha_1P+q_\perp$). Using Eq.~(\ref{bswf}), we can integrate out $q^0$ and get
\begin{equation}
\begin{aligned}\label{amp2}
{\mathcal M} &= -i\int\frac{d^3\vec{q}}{(2\pi)^3}{\rm{Tr}}[\varphi(\vec{q})(\slashed\epsilon_2\frac{1}{\slashed p_1-\slashed k_1-(m_1-i\epsilon)}\slashed\epsilon_1+\slashed\epsilon_1\frac{1}{\slashed p_1-\slashed k_2-(m_1-i\epsilon)}\slashed\epsilon_2)]\\
&= \frac{4}{M} \epsilon_{\alpha\beta\sigma\delta}\epsilon_1^\alpha\epsilon_2^\beta P^\delta\epsilon_{\mu\nu}k_1^\sigma \int\frac{d^3q_\perp}{(2\pi)^3}(\frac{1}{(\vec q+\vec k_1)^2+m_1^2} + \frac{1}{(\vec q-\vec k_1)^2+m_1^2})q_\perp^\mu q_\perp^\nu f_2 .
\end{aligned}
\end{equation}
One notices that this result has a similar structure as in Eq.~(\ref{m1integral}). Actually, if we make the approximation $m_1\sim\omega_1\sim\alpha_1 M$ and $f_1\sim f_2$, Eq.~(\ref{m1integral}) will reduce to Eq.~(\ref{amp2}).

In Table~\ref{gamma} and \ref{gluon}, we give the two-gamma and two-gluon decay
widths for the first five $2^{-+}$ heavy quarkonia. The results of two methods are in good agreement with each other, while the one with $p_1^0 = \frac{M}{2}$ approximation is larger. As the principal quantum number increases, the decay width first increases and then decreases for charmonium,
while it increases all the time for bottomonium. This can be
understood as follows. First, the overlap integral $E$ ($E^\prime$)  in Eq.~(\ref{es})
decreases as the principal quantum number increases
(for bottomonium, $E$ first increases, then decreases). This is
because the cancellation is more severe if the
wave function has more nodes. Second, there is also a factor $M^7$ which is
increasing all the time. The final result comes from combining
these two effects.

We present two-gamma decay widths for the first two charmonium and bottomonium states with different models in Table~\ref{compare1}. Our result is close to that of Ref.~\cite{AB} and~\cite{Munz}, which are based on the semi-relativistic and relativistic formalism, respectively. Ref.~\cite{Novikov} gives the non-relativistic result which is about 4 times of ours. With the two-body Dirac equation (TBDE) method, Ref.~\cite{Crater} gets a much larger result. In Table~\ref{compare2}, two-gluon decay widths for the ground state with different models are given. Ref.~\cite{Volk} use the known values of charmonia decay widths into different channels and the non-relativistic formulas to get a result which is about 1.7 times of ours. Ref.~\cite{ELQ, GB, Novikov} uses the non-relativistic method. The deviation of the results comes from the use of different parameters. Using the NRQCD method, Ref.~\cite{Chao1} gets the leading order result which is about 4 times of ours.

Our result shows great relativistic corrections exist in the annihilation processes of D-wave
quarkonia. This can be understood from the behavior of the wave
function in momentum space. In Fig.~\ref{wf}, we plot the ground-state wave
functions of the $0^{-+}$ and $2^{-+}$ charmonia. For $1^1D_2(c\bar c)$, one can see
the wave function reaches its maximum when $q$ takes a relatively large
value ($\sim0.7$ GeV), which means the
relativistic corrections will be significant. This can be compared
with the $\eta_c$ (S-wave) case. There the dominant contribution
of the wave function comes from the range where $q$ is small, but
even so the relativistic corrections is about $30\sim40\%$.

For the state below $DD^\ast$ or $BB^\ast$ threshold, we can
estimate its total decay width by adding the widths of different
channels together as mentioned previously. For $1{^1D_2(c\bar c)}$,
we can get $\Gamma \approx \Gamma_{gg}+\Gamma_{\gamma
h_c}+\Gamma_{\gamma J/\psi}$ = 432 keV, where we have used the
results of $\Gamma_{\gamma h_c}$ and $\Gamma_{\gamma J/\psi}$ in
Ref.~\cite{wang2}. With this total decay width we can estimate the
branching ratio of the two-gamma channel: Br$(1{^1D_2}(c\bar
c)\rightarrow\gamma\gamma)=2.7\times 10^{-5}$. This is close to
that of $\eta_c$, while is one order of magnitude  less than that
of $\chi_{c0}$ and $\chi_{c2}$~\cite{PDG}. The golden channel to find the $1{^1D_2(c\bar c)}$ is its decay to $\gamma h_c$ since its branching ratio is about $90\%$.

In summary, by using the instantaneous BS method, we have calculated
two-photon and two-gluon annihilation processes of charmonium
and bottomonium states. It shows that the decay width of the charmonium
increases first and then decreases, while the decay width keeps increasing for the bottomonium case. The relativistic
corrections are very large, which implies one has to include them. Although the two methods
we used in this paper give results very close to each other, the one without the $p_1^0 = p_2^0 = \frac{M}{2}$ approximation is more justifiable.

\section{Acknowledgments}

This work was supported in part by the National Natural Science
Foundation of China (NSFC) under Grant No.~11175051.

\end{document}